\documentclass{iaus}
\usepackage{graphicx}
\usepackage{amsmath} 
\newcommand{\bea}{\begin{eqnarray}} 
\newcommand{\eea}{\end{eqnarray}} 
\newcommand{\nn}{\nonumber} 
\newcommand{\vek}[1]{\boldsymbol{#1}}

\newcommand{\aap}{Astronomy \& Astrophysics~}
\newcommand{\apj}{Astrophysical Journal~}
\newcommand{\aj}{Astronomical Journal~}

\newcommand{\plotone}[1]{\includegraphics[width=5.0in,height=3.in]{#1}}
\title[Black holes in AGN] 
{Black holes in Active Galactic Nuclei}
\author[Valtonen et al.]{M.~J.Valtonen$^{1}$, S.~Mikkola$^{1}$, 
D.~Merritt$^2$, 
A.~Gopakumar $^{3,8}$, 
H.~J.~Lehto$^{1}$, T.~Hyv\"onen$^{1}$, 
H.~Rampadarath$^{4,5}$, R.~Saunders$^{6}$, 
M.~Basta$^{7}$ and R.~Hudec$^{7,9}$} 
\affiliation{$^1$Tuorla Observatory, Department of Physics and Astronomy, University of Turku, 
    21500 Piikki\"o, Finland\\ \noindent
$^2$ Centre for Computational Relativity and Gravitation, Rochester Institute of Technology, 
78 Lomb Memorial Drive, Rochester, NY 14623, USA \\ 
$^3$Tata Institute of Fundamental Research, Mumbai 400005, India\\
$^4$Joint Institute for VLBI in Europe (JIVE), Postbus 2, 7990 AA Dwingeloo, The
Netherlands\\
$^5$Leiden Observatory, Leiden University, P.O. Box 9513, NL-2300 RA  Leiden, The Netherlands\\ 
$^6$ Department of Physics, University of the West Indies, St. Augustine, Trinidad \& Tobago\\
$^7$ Astronomical Institute, Academy of Sciences, Fricova 298, 25165 Ondrejov, Czech Republic\\
$^8$Theoretisch-Physikalisches Institut,
Friedrich-Schiller-Universit\"at Jena,\\
Max-Wien-Platz 1, 07743 Jena, Germany\\
$^9$ Czech Technical University in Prague, Faculty of Electrical
Engineering, Technická 2, 166 27 Praha 6, Czech Republic}

\pubyear{2008}
\volume{261}  
\pagerange{1--10}
\jname{Relativity in Fundamental Astoronomy}
\editors{S.A.Klioner, K.P.Seidelman \& M.H. Soffel, eds.}
\begin{document}
\maketitle
\begin{abstract}
Supermassive black holes are common in centers of galaxies. Among the active
galaxies, quasars are the most extreme, and their black hole masses range as
high as to $6\cdot 10^{10} M_\odot$. Binary black holes are of special interest
but so far OJ287 is the only confirmed case with known orbital elements. In
OJ287, the binary nature is confirmed by periodic radiation pulses. The period
is twelve years with two pulses per period. The last four pulses have been
correctly predicted with the accuracy of few weeks, the latest in 2007 with the
accuracy of one day. This accuracy is high enough that one may test the higher
order terms in the Post Newtonian approximation to General Relativity. The
precession rate per period is  $39^\circ\!.1 \pm 0^\circ\!.1$, by far the largest rate in any
known binary, and the  $(1.83\pm 0.01)\cdot 10^{10} M_\odot$ primary is among
the dozen biggest black holes known. We will discuss the various Post Newtonian
terms and their effect on the orbit solution.
 The over 100 year data base of optical variations in OJ287 puts limits
on these terms and thus tests the ability of Einstein's General
Relativity to describe, for the first time,  dynamic binary black hole
spacetime in the strong field regime.
The quadrupole-moment contributions to the equations of motion
allows us to constrain the `no-hair' parameter to be $1.0\:\pm\:0.3$
which supports the black hole no-hair theorem within the achievable
precision.
\end{abstract}


\firstsection

\section{Introduction} 
Centers of galaxies typically host dark massive bodies which are thought to be
black holes. The black hole concept plays a particularly strong role in
theories of quasars and other active galactic nuclei. Some models use the 
assumed black hole properties of the dark bodies, such as the frame dragging
around a spinning black hole (e.g. Camenzind 1989), but on the whole the 
activity is largely explained by accretion onto a central heavy body, as was
first pointed out by Salpeter (1964) and Zeldovich (1964). Theories of 
accretion disks have been developed, in particular the $\alpha$ disk theory
of Shakura and Sunyaev (1973) and its extension to magnetic disks by Sakimoto
and Corotini (1981) which describe well the origin of activity around massive
central bodies, whatever the exact nature of these bodies may be.

In order to
prove that the central body is actually a black hole we have to probe the
gravitational field around it. One of the most important characteristics of a
black hole is that it must satisfy the so called no-hair theorem or theorems
(Israel 1967, 1968, Carter 1970, Hawking 1971, 1972; see Misner, Thorne and 
Wheeler 1973). A  practical test was suggested by Thorne (1980) and Thorne,
Price and Macdonald (1986). In this test the quadrupole moment $Q$ of the
spinning body is measured. If the spin of the body is $S$ and its mass is $M$,
we determine the value of $q$ in
\begin{eqnarray}
Q = -q \frac{S^2}{Mc^2}.
\end{eqnarray}
For black holes $q=1$, for neutron stars and other possible bosonic structures
$q > 2$.
This is an important test for stellar mass bodies where stable configurations
of various kinds may exist (Wex and Kopeikin 1999), but it is also a prime test
for the supermassive black hole concept in Active Galactic Nuclei and even in
our Galactic Center (Will 2008).

\section{OJ287}

BL Lacertae object OJ287 is known to have a quasiperiodic pattern of outbursts
at 12 year intervals
(\cite[Sillanp\"a\"a et al.(1988)]{sil88},\cite[Valtonen et al.(2008c)]{val08c}). 
Further, the light curve has a double peak structure, with the two peaks
separated by 1 - 2 years. The available information is that the radiation at the peaks is
thermal bremsstrahlung radiation, in contrast to the non-thermal synchrotron emission at
'normal' times. The origin of the thermal bremsstrahlung is thought to be an impact of 
a secondary black hole on the accretion disc of the primary (\cite[Lehto \& Valtonen (1996)]{leh96}). 
The model has been successful in predicting future outbursts: 
the predictions for the beginning of 1994, 1995 and 2005 outbursts 
were correct within a few weeks. In addition to timing the impacts on the accretion disk, the 
model includes delays of outbursts relative to the disk crossing. The outburst begins when
a bubble of gas torn off the accretion disk becomes transparent at optical wavelengths. 
For the 2005 outburst, 
it also was necessary to consider the bending of the accretion disk caused by 
the secondary (\cite[Sundelius et al. (1996)]{sun96},\cite[Sundelius et al. (1997)]{sun97}); 
when combined with the earlier model 
for the radiation burst delay \cite[Lehto \& Valtonen (1996)]{leh96} the beginning of the 2005 outburst 
was expected at 2005.74. This is very close to the actual starting time of the outburst
(\cite[Valtonen et al.(2006a)]{val06a}, \cite[Valtonen et al.(2008a)]{val08a}). It confirmed the need for relativistic 
precession since without the precession the outburst would have been a year later. 
Finally, the gravitational radiation energy loss was included in the prediction 
for the next outburst in September 2007 (\cite[Valtonen (2007)]{val07}, \cite[Valtonen et al.(2008a)]{val08a}). 
Observations confirmed the correctness of the prediction within the accuracy of one day
\cite[Valtonen et al.(2008c)]{val08c}. 
The model that does not incorporate the gravitational radiation reaction effect is clearly
not tenable since 
it predicted the outburst three weeks too late. 
The probability of so many major outbursts happening at the predicted times by chance is
negligible. 

Recently, models have been developed which include the spin of the primary black hole
(Valtonen et al. 2009). Here we will review this work and discuss its general implications.

\section{Observations} 
\begin{figure}[t] 
\plotone{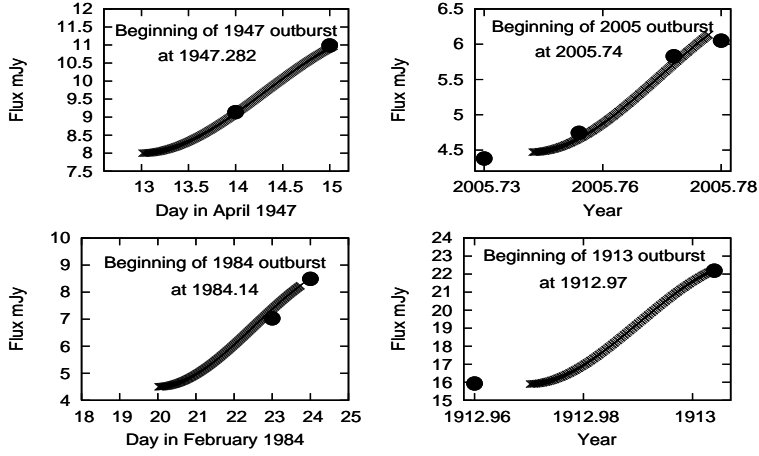}
\caption{The observation of the brightness of OJ287 in four different seasons: 1912/3, 1947, 1984 and 2005. The heavy curve is a fit through the observed points (dots). Each panel labels the deduced start of the outburst.\label{fig1}} 
\end{figure} 
The model requires accurate timing of the outbursts. The outburst is thought to arise 
from a hot bubble of gas which has cooled to a point where it becomes suddenly 
optically transparent. When such a bubble is viewed from a distance, the emission is seen to
grow in a specific way as the observational front advances into the bubble. The size of the
bubble, and thus also the rate of development of the outburst light curve, is a known function
of distance from the center of the accretion disk. In Figures 1 and 2 we display theoretical light curves
for different outbursts and overlay them with observations. The labels inside the panels tell the timing which is related to the image.    
\begin{figure}[ht] 
\plotone{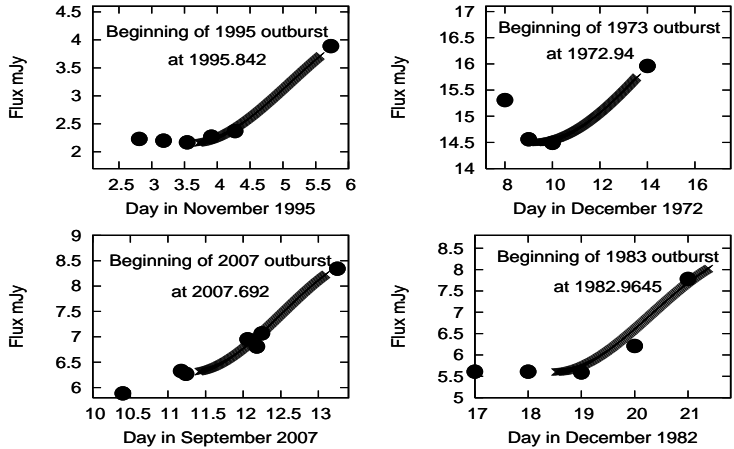}
\caption{The observation of the brightness of OJ287 in four different seasons: 1972/3, 1982/3, 1995 and 2007. See caption of Fig. 1. 
\label{fig2}}
\end{figure} 
\section{PN-accurate orbital description} 

  We invoke the 2.5PN-accurate orbital dynamics that includes the leading order 
general relativistic, classical spin-orbit and radiation reaction effects 
for describing the temporal evolution of a binary black hole (\cite[Kidder (1995)]{kid95}). 

The 2.5PN-accurate equations of motion can be written schematically as 
\begin{eqnarray} 
\ddot { {\vek x}} \equiv 
\frac{d^2  {\vek x}} { dt^2} &=& 
\ddot { {\vek x}}_{0} + \ddot { {\vek x}}_{1PN} \nonumber 
+ \ddot { {\vek x}}_{SO}+\ddot { {\vek x }}_{Q}\\ 
&& + \ddot { {\vek x}}_{2PN} +  \ddot { {\vek x}}_{2.5PN} \,,  
\end{eqnarray} 
where ${\vek x} = {\vek x}_1 - {\vek x}_2 $ stands for the 
center-of-mass 
relative separation vector between the black holes with masses $m_1$ and $m_2$ and 
$  \ddot { {\vek x}}_{0}  $ represents the Newtonian acceleration given by  
$ \ddot { {\vek x}}_{0} = -\frac{ G\, m}{ r^3 } \, {\vek x} $; $m= m_1 + m_2$ and $ r = | {\vek x} |$. 
The PN contributions occurring at the conservative 1PN, 2PN and the reactive 2.5PN orders, denoted 
by $\ddot { {\vek x}}_{1PN}$, $\ddot { {\vek x}}_{2PN}$ and 
$\ddot { {\vek x}}_{2.5PN}$ respectively, are non-spin by nature. The explicit expressions for these 
contributions, suitable for describing the 
binary black hole dynamics, were derived for the first time in the harmonic gauge 
(\cite[Damour (1982)]{TD}). These are well known and they are not repeated here.

The leading order spin-orbit contributions to $ \ddot {{\vek x} }$, 
appearing at 1.5PN order (\cite[Barker \& O'Connell (1975)]{bar79}), reads 
\begin{eqnarray} 
& \ddot { {\vek x}}_{SO} = \frac{ G^2 m^2}{c^3 r^3}
\left ( \frac{ 1 + \sqrt{1 -4\,\eta} }{4} \right )\chi\cdot &\\ & \nn 
\,  
\biggl \{ \biggl [ 12 \, \left [ {\vek s}_1 \cdot (  {\vek n} \times {\vek v} ) \right ] \biggr ]\, 
 {\vek n} 
+ \biggl [ \left ( 9 + 3\, \sqrt {1 - 4\, \eta} \right ) \, \dot r \biggr ] 
\left (  {\vek n} \times {\vek s}_1 \right ) 
\nn 
- \biggl [ 
7 +  \sqrt {1 - 4\, \eta} 
\biggr ] 
\left ( {\vek v} \times {\vek s}_1 \right ) 
\biggr \}\,,\nn 
\end{eqnarray} 
where the vectors $ {\vek n}$ and ${\vek v}$ are defined to be 
$  {\vek n} \equiv {\vek x}/r $ and 
$ {\vek v} \equiv d {\vek x}/dt $, respectively, while 
$ \dot r \equiv dr/dt =  {\vek n} \cdot {\vek v}$, 
$ v \equiv | {\vek v} |$ and the symmetric mass ratio $\eta = m_1\, m_2/m^2$. 
The Kerr parameter $\chi$ and the unit vector 
${\vek s}_1$ define the spin of the primary black hole by the relation 
${\vek S}_1 = G\, m_1^2 \, \chi\, {\vek s}_1/c$ 
and $\chi$ is allowed to take values between $0$ and $1$ in general relativity. 
Further, the above expression for $\ddot { {\vek x}}_{SO}$ implies that the  
covariant spin supplementary condition is employed to define the center-of-mass world line of 
the spinning compact object in the underlying PN computation (\cite[Kidder (1995)]{kid95}). 
Finally, the quadrupole-monopole interaction term $\ddot { {\vek x}}_Q $, entering at the 2PN order (\cite[Barker \& O'Connell (1975)]{bar79}), reads
\begin{eqnarray} 
\ddot { {\vek x}}_Q & =- q \, \chi^2\, 
\frac{3\, G^3\, m_1^2 m}{2\, c^4\, r^4}
\, \biggl \{ 
\biggl [ 5(\vek n\cdot \vek s_1)^2 
-1 \biggr ] {\vek n}
\nn
-2(\vek n\cdot \vek s_1) {\vek s_1} \biggr \}, 
\end{eqnarray} 
where the parameter $q$, whose value is $1$ in general relativity, is introduced to test the black hole `no-hair'   
theorem.
The precessional motion for the 
primary black hole spin is dominated by 
the leading order general relativistic spin-orbit coupling 
and the relevant equation reads 
\begin{eqnarray}\nn
\frac{d {\vek s}_1}{dt} = {\vek \Omega} \times {\vek s}_1 \,,~~ ~ ~ ~ ~ ~ ~ ~ ~ ~ ~ ~ ~ ~  ~  ~  ~  ~  ~  ~  ~  ~   \\ 
{\vek \Omega} = \left ( \frac{G\,m\, \eta}{2c^2\, r^2} \right ) \,  
\biggl ( \frac{ 7 + \sqrt {1 - 4\, \eta} }{ 1 + \sqrt {1 - 4\, \eta} } 
\biggr ) 
\, \, \left (  {\vek n} \times  {\vek v} \right ) 
.\end{eqnarray} 
It should be noted that the precessional equation for the unit spin vector ${\vek s}_1$ enters the 
binary dynamics at 1PN order, while the spin contribution enters 
$ \ddot {{\vek x} }$ 
at the 1.5PN order. 

  The main consequence of including the leading order spin-orbit interactions to the dynamics 
of a binary black hole is that it forces both the binary orbit and the orbital plane to precess. Moreover, 
the orbital angular momentum vector, characterising the orbital plane, precesses around 
the spin of the primary in such a way that the angle between the orbital plane and 
the spin vector ${\vek s}_1 $ remains almost constant (roughly within $\pm 0.^\circ\!5$ in our model). 
The spin-vector itself precesses drawing a cone with an opening angle of about 12 degrees. 


\begin{figure}[t] 
\plotone{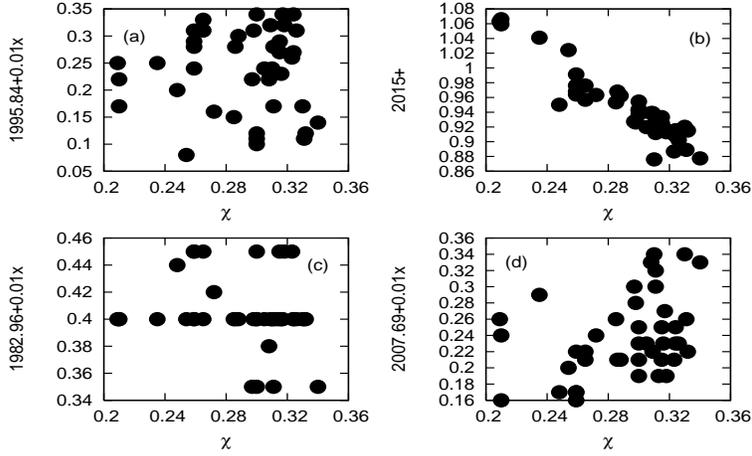}
\caption{The dots represent solutions with different values of $\chi$ (horizontal axis) and outburst time (vertical axis). On the y-axis of panel (a) the points scatter around 1995.842, in panel (c) around 1982.964 and in panel (d) around 2007.692. In contrast, panel (b) shows a strong correlation between the outburst time in 2015 and $\chi$.\label{fig3} 
  } 
\end{figure} 

  \begin{figure}[t] 
\plotone{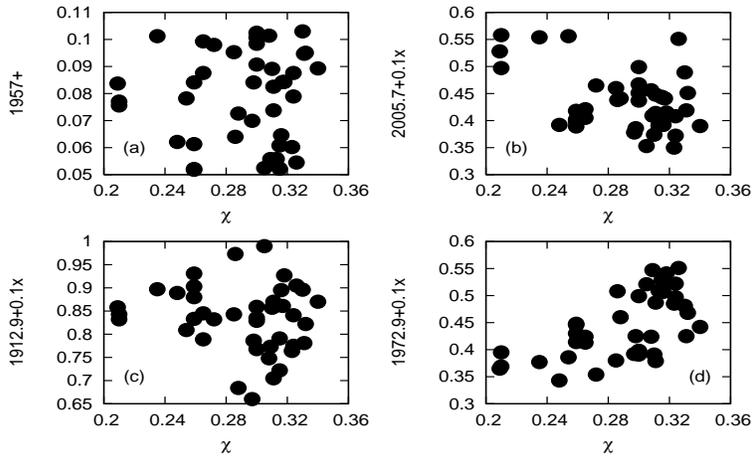}
  \caption{The dots represent solutions with different values of $\chi$ (horizontal axis) and outburst time (vertical axis). On the y-axis of panel (a) the points scatter around 1957.08, in panel (b) around 2005.745, in panel (c) around 1912.98 and in panel (d) around 1972.945.\label{fig4}
} 
\end{figure}

The precessional period for both the orbital plane and the spin of the binary
is about 2400 years.  The orbital inclination 
relative to the plane of symmetry of the spinning black hole ( and presumably relative to the 
accretion disk of the primary as well) is taken to be 90 degrees.

\section{Timing experiments}

In previous work (e.g. Valtonen 2007) we do the timing experiments using the 1913, 1947, 1973, 1983, 1984 and 2005 outbursts as fixed points. We cannot determine the Kerr parameter $\chi$ and the no-hair parameter $q$ without considering more fixed points.
\begin{figure}[t]  
\plotone{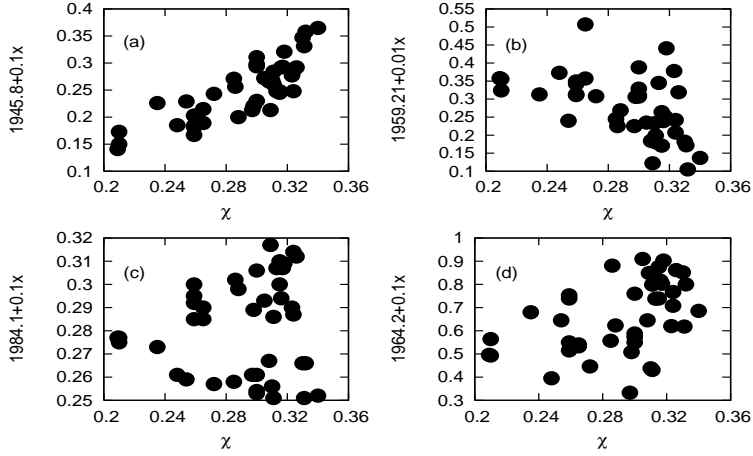}
\caption{The dots represent solutions with different values of $\chi$ (horizontal axis) and outburst time (vertical axis). On the y-axis of panel (b) the points scatter around 1959.213, in panel (c) around 1984.129, and in panel (d) around 1964.27. Panel (a) displays a strong correlation between the timing of the 1945 outburst and $\chi$. There are not enough data at present to verify the timings of 1959, 1964 and 1945 outbursts, while the 1984 outburst is used as a fixed point in our solutions. If new data points are found from years 1945 - 1964, the model can be further refined.\label{fig5}}
\end{figure}  
Here we also make use of the 1995 and 2007 outbursts which have been observed with high time
resolution (see Figure~\ref{fig2}) and they are therefore suitable for further refinement of the model. In addition, the newly discovered 1957 outburst is taken as another fixed point
(\cite[Valtonen et al. (2006b)]{val06b}, \cite[Rampadarath et al. (2007)]{ram07}). The observed outburst times are listed in 
Table \ref{outburst}.

With this full set of outbursts no solutions were found unless the parameter $\chi$ is in the range 0.2 - 0.36. The
majority of solutions cluster around $\chi=0.29$, with a one standard deviation of about 0.04.
 \begin{figure}[t] 
\plotone{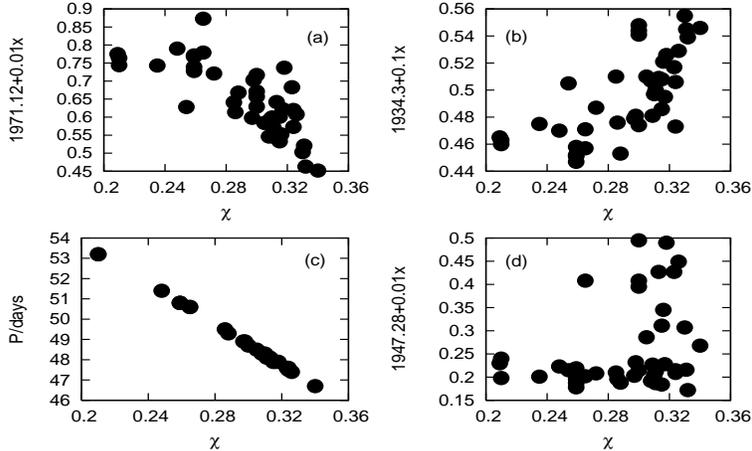}
\caption{The dots represent solutions with different values of 
$\chi$ (horizontal axis) and outburst time (vertical axis). 
On the y-axis of panel (a) the points scatter around 1971.126, 
in panel (b) around 1934.35 and in panel (d) around 1947.283.
In contrast, panel (c) shows a dependence of the half-period of 
the innermost stable orbit on $\chi$.
The observed half-period is between 43 and 49 days, 
implying that $\chi$ is likely to be at least 0.29. 
The 1947 outburst is one of our fixed points, while there are not enough data to
verify 1934 and 1971 outbursts at present.\label{fig6}}
\end{figure}
\begin{table}[t] 
\caption{Outburst times with  estimated uncertainties\label{outburst}. 
These are starting times of the outbursts, normalised to the event of 1982.964.} 
\begin{center}
\begin{tabular}{lr} 
1912.980& $\pm$  0.020\\ 
1947.283& $\pm$  0.002\\ 
1957.080& $\pm$  0.030\\ 
1972.945& $\pm$  0.012\\ 
1984.130&$\pm$   0.005\\ 
1995.842&$\pm$   0.0015\\ 
2005.745& $\pm$  0.012\\ 
2007.692&$\pm$   0.0015\\ 
\end{tabular} 
\end{center}
\end{table} 
The timing experiments give a unique solution for the system parameters: \:\: 
precession in the orbit plane per period $\Delta \phi$, 
masses $m_1$ and $m_2$ of the two black holes, 
spin parameter $\chi$, initial phase $\phi_0$, initial apocenter eccentricity $e_0 $, 
`no-hair' parameter $q$ and 
time-delay parameter $t_d$. The results are shown in Table \ref{solutiontable}. 

Here the precession rate is defined as the average change of the apocenter phase angle over the last 150 yrs.
The time delay parameter depends on the structure of the accretion disk, and obtains 
the value of unity in the model of 
\cite[Lehto \& Valtonen (1996)]{leh96}.  
Its value is related to the thickness of the accretion disk, as explained in \cite[Valtonen et al. (2006b)]{val06b}. 
\begin{table}[h] 
\caption{Solution parameters.\label{solutiontable}} 
\begin{center}
\begin{tabular}{l|r} 
$\Delta \phi^\circ$  & $39.1 \pm 0.1$\\ 
$m_1$ & $(1.83\pm 0.01)\cdot 10^{10} M_\odot$ \\ 
$m_2$ & $(1.4\pm 0.1)\cdot 10^8 M_\odot$\\ 
$\chi$ &  $0.28\pm 0.08$\\ 
$\phi_0$ & $ 56^\circ\!.5\ \pm 1^\circ\!.2$\\ 
$e_0 $ & $0.6584\pm 0.001$\\ 
$q$ & $1.0\pm 0.3$\\ 
$t_d$& $0.75\pm 0.04$\\ 
\end{tabular} 
\end{center}
\end{table} 
Figures 3-6 show the distribution points, each representing a solution, in four panels per figure. 
Each panel displays the outburst time as a function of $\chi$. 
In Figure 3 the panels (a), (c) and (d) are essentially scatter diagrams, 
demonstrating that the solutions cover the allowed range rather well. 
On the contrary, the panel (b) shows a strong correlation between the time 
of the 2015 outburst and $\chi$. The range of possible outburst times extends 
from early November if $\chi=0.36$ to late January 2016 if $\chi=0.2$.  

\section{More OJ287's?} 

How unique is OJ287? Can we hope to observe more similar systems? OJ287 is highly variable mostly due to strong beaming. At its faintest it goes to $m_B=18$ which may be taken as its intrinsic (unbeamed) brightness. There are about $2\cdot 10^{4}$ quasars in the sky brighter than this magnitude limit (Arp 1981).
Many of these quasars host binary black holes, perhaps as many as 
50\% (Comerford et al. 2009). Thus potentially there are about $10^{4}$ bright binary quasars in the sky to be discovered.
But are they all similar to OJ287? OJ287 is in the last stages of inspiral, with only $10^4$ yr left out of its potential $10^7$ yr lifetime (Volonteri et al. 2009). Thus the number of bright short period binaries which could be discovered by techniques similar to the discovery of OJ287 is only of the order of ten over the whole sky. 
Another property of OJ287 which may be relevant to the discovery rate, 
is its mass at the upper end of black hole mass range. Even though a high mass generally 
means high luminosity and sampling of large volumes of space 
(observed numbers increasing roughly as $M^{1.5}$ for luminosity being a 
constant fraction of Eddington luminosity), the frequency function of black 
hole mass falls with increasing mass. The power law slope at the high mass end is about -2.6, 
i.e. the probability density for mass being greater than $M$ has a power law index of about -1.6 
(Vestergaard and Osmer 2009). Thus the effect of frequency function is to compensate 
for the volume effect (within the uncertainties). 
For a binary system with a constant period, there is also another factor. 
The remaining lifetime of the binary goes down as $M^{-2/3}$. 
The most likely value for a mass from this distribution is one quarter of the upper limit 
(which is of the order of $6\cdot 10^{10} M_\odot$, Vestergaard et al. 2008). 
It is not too far from our measured mass value.
The main reason for the discovery of the OJ287 binary in a sample of about 50 quasars 
studied by the Tuorla monitoring group is the fact that it is a blazar. Because of this, it was a prime target for variability studies. The double peak optical outbursts would have easily gone unnoticed if it did not attract attention by its variable jet brightness which is seen all the time. Therefore one should extend the periodicity searches to quasars which are not highly variable, in hopes of detecting occasional double peaked outbursts.

From the measured frequency of over $10^{10}M_\odot$ black holes in quasars one easily estimates that there must be a dozen or more such quasars
in the local volume encompassing OJ287. Since their jets would not typically point toward us, they would only appear bright during the short
intervals of disk impacts ; it would require a major program of historical plate analysis and observational monotoring 
to catch these potentially OJ287-type quasars in action.

\section{Parameter values}
We have already pointed out that the primary mass $m_1$ = $(1.83\pm 0.01)\cdot 10^{10} M_\odot$ 
is quite consistent with what might be expected on general grounds. 
What about the secondary mass? Its value was calculated by Lehto and Valtonen 
(1996) as $m_2\approx 1.23\cdot 10^{8} M_\odot$, 
updated to today's Hubble constant, while Sundelius et al. (1997) preferred a 30\% greater value. 
These estimates are based on the astrophysics of the disk impacts and on the amount of 
radiation produced in these events, and thus they are totally independent of the orbit model. 
These estimates agree well with our more exact value $m_2 = (1.4\pm 0.1)\cdot 10^8 M_\odot$.

The eccentricity of the orbit, if defined by using pericenter and apocenter distances as for a Keplerian orbit, 
is $e=0.7$. The initial eccentricity at the beginning of the final inspiral, say, 
$10^5$ yr prior to merger, must have been about $e=0.9$, a reasonable value at the initial 
stage of merging black holes (Aarseth 2008). Then by today the eccentricity would 
have evolved to its current value.

The parameter $t_d$ gives the ratio of $\alpha_g$ to the accretion rate in Eddington units. 
The value $t_d=0.75\pm 0.04$ implies $14\pm0.5$ for this ratio. 
The mass accretion rate may be about   $0.005$ of the Eddington rate (Bassani et al. 1983) which 
gives $\alpha_g=0.07\pm0.03$, a reasonable value.

We infer that the primary black hole 
should spin approximately at one quarter of the maximum spin rate allowed in general relativity. There is an additional observation which supports this spin value. OJ287 has a
basic 46$\pm$3 day periodicity (\cite[Wu et al.(2006)]{Wu06}), which may be related to the innermost stable orbit in the accretion disk. 
However, since we presumably
observe the accretion disk almost face on, and there is an $m=2$ mode wave disturbance in the disk, this is likely to refer
to one half of the period. Considering also the redshift of the system, and the
primary mass value given in Table 2 this corresponds to the spin of $\chi= 0.35 \pm0.06$ (\cite[McClintock et al. (2006)]{McClintocketal}). The uncertainty of 0.06 units is related to the
width of a trough in the structure function of the flux variations. 
For comparison, it has been estimated that $\chi\sim 0.5$ for the black hole in the Galactic center (Genzel et al. 2003).

The values of $q$ cluster around $q=1.0$ with a standard deviation of $0.1$ and maximum range $\pm0.3$. This
is the first indication that the 'no-hair' theorem is valid. Even though there are no other known stable configurations than a black hole in the supermassive scale, it is still interesting that our result converges at the proper value for general relativity.


\begin{thebibliography}{99} 
\bibitem[Aarseth (2008)]{aar08} Aarseth, S.J. 2008, in Dynamical Evolution of Dense Stellar Systems, Proceedings of IAU Symp. 246, Eds. E. Vesperini, M. Giersz \& A. Sills, Cambridge Univ. Press, Cambridge, p. 437
\bibitem[Arp (1981)]{arp81} Arp, H. 1981, \apj 250, 31 
\bibitem[Barker \& O'Connell (1975)]{bar79} Barker, B. M. \& O'Connell, R. F.  1975, PhRvD, 12, 329
\bibitem[Bassani et al. (1983)]{bas83} Bassani,L., Dean, A.J. \& Sembay, S., \aap, 125, 52
\bibitem[Camenzind (1989)]{cam89} Camenzind, M. 1989, in Accretion Disks and Magnetic Fields in Astrophysics, Ed. G. Belvedere, Kluwer, Dordrecht, p. 129
\bibitem[Carter (1970)]{car70}  Carter, B. 1970, Phys.Rev.Lett. 26, 331
\bibitem[Comerford et.al (2009)]{com09}  Comerford, J.M., Gerke, B.F., Newman, J.A., Davis, M., Yan, R., Cooper, M.C., Faber, S.M., Koo, D.C., Coil, A.L., Rosario, D.J. \& Dutton, A.A. 2009, \apj, 698, 956

\bibitem[Damour (1982)]{TD}  Damour, T. 1982, C.R. Acad. Sci. Paris 294, (II), 1355. 
\bibitem[Genzel et al. (2003)]{Genzeletal}Genzel, R., Schodel, R., Ott, T., Eckart, A., Alexander, T., Lacombe, F., Rouan, D. and Aschenbach, B. 2003, Nature 425, 934 
\bibitem[Hawking (1971)]{haw71} Hawking, S.W. 1971, Phys.Rev.Lett. 26, 1344
\bibitem[Hawking (1972)]{haw72} Hawking, S.W. 1972, Commun.Math.Phys. 25, 152
\bibitem[Israel (1967)]{isr67} Israel, W. 1967, Phys.Rev. 164, 1776
\bibitem[Israel (1968)]{isr68} Israel, W. 1968, Commun. Math.Phys. 8, 245

\bibitem[Kidder (1995)]{kid95} Kidder, L. E. 1995, PhRvD, 52, 821 
\bibitem[Lehto \& Valtonen (1996)]{leh96} Lehto, H.J., \& Valtonen, M.J. 1996, \apj, 460, 207 
\bibitem[Misner et al. (1973)]{mis73} Misner, C.W., Thorne, K.S. \& Wheeler, J.A. 1973, Gravitation, W.H.Freeman \& Co, New York, p. 876

\bibitem[Rampadarath et al. (2007)]{ram07} Rampadarath, H., Valtonen, M. J., \& Saunders, R. The Central Engine of Active Galactic Nuclei, Eds. Luis C. Ho \& Jian-Min Wang, ASP Conf. Ser., 373, 243 
\bibitem[Sakimoto \& Corotini (1981)]{sak81} Sakimoto, P.J. \& Corotini, F.V. 1981, \apj, 247, 19
\bibitem[Salpeter (1964)]{sal64} Salpeter, E.E. 1964, \apj, 140, 796
\bibitem[Shakura \& Sunyaev (1973)]{sha73} Shakura, N.I. \& Sunyaev, R.A. 1973, \aap, 24, 337
\bibitem[Sillanp\"a\"a et al.(1988)]{sil88} Sillanp\"a\"a, A., Haarala, S., Valtonen, M.J., Sundelius, B. \& Byrd,     G.G.    1988, \apj, 325, 628 
\bibitem[Sillanp\"a\"a et al.(1996a)]{sil96a} Sillanp\"a\"a et al.  1996a, \aap, 305, L17 
\bibitem[Sillanp\"a\"a et al.(1996b)]{sil96b} Sillanp\"a\"a et al.  1996b, \aap, 315, L13 
\bibitem[Sundelius et al. (1996)]{sun96} Sundelius, B., Wahde, M., Lehto, H.J. \& Valtonen, M.J., 1996, Blazar Continuum Variability, ASP Conf. Ser., 110, 99 
\bibitem[Sundelius et al. (1997)]{sun97} Sundelius, B., Wahde, M., Lehto, H.J. \& Valtonen, M.J., 1997, \apj, 484, 180 
\bibitem[Thorne (1980)]{tho80} Thorne, K.S. 1980, Rev.Mod.Phys. D, 31, 1815
\bibitem[Thorne et al. (1986)]{tho86} Thorne, K.S., Price, R.M. \& Macdonald, D.A. 1986, in Black Holes: The Membrane Paradigm, Yale Univ. Press, New Haven

\bibitem[Valtonen \& Lehto (1997)]{val97}  Valtonen, M.J. \& Lehto, H.J. 1997, \apj, 481, L5 
\bibitem[Valtonen et al.(2006a)]{val06a}  Valtonen, M.J. et al. 2006a, \apj, 643, L9 
\bibitem[Valtonen et al. (2006b)]{val06b}  Valtonen, M.J. et al. 2006b, \apj, 646, 36 
\bibitem[Valtonen (2007)]{val07}  Valtonen, M.J. 2007, \apj, 659, 1074 
\bibitem[Valtonen et al.(2008a)]{val08a}  Valtonen, M.J., Kidger, M., Lehto, H. \& Poyner, G., 2008b, \aap, 477, 407 
\bibitem[Valtonen (2008)]{val08b}  Valtonen, M.J. 2008, RevMexA\&Ap, 32, 22 
\bibitem[Valtonen et al.(2008c)]{val08c}  Valtonen, M.J. et al., 2008c, Nature, 452, 851 
\bibitem[Valtonen et al.(2009)]{val09} Valtonen, M.J., Mikkola,S., Merritt, D., Gopakumar, A., Lehto, H.J., Hyv\"onen, T., Rampadarath, H., Saunders, R., Basta, M. \& Hudec, R. 2009, submited
\bibitem[Vestergaard et al. (2008)]{ves08} Vestergaard, M., Fan, X., Tremonti, C.A., Osmer, P.O. \& Richards, G.T. 2008, \apj, 674, L1
\bibitem[Vestergaard \& Osmer (2009)]{ves09} Vestergaard, M., \& Osmer, P.S., arXiv:0904.3348v1[astro-ph]
\bibitem[Volonteri et al. (2009)]{vol09} Volonteri, M., Miller, J.M. \& Dotti, M. 2009, arXiv:0903.3947v1[astro-ph]
\bibitem[Wex \& Kopeikin (1999)]{wex99} Wex, N. \& Kopeikin, S.M. 1999, \apj, 514, 388

\bibitem[Will(2008)]{Will2008} Will, C.~M.\ 2008, Astrophysical Journal 674, L25-L28. 
\bibitem[Wu et al.(2006)]{Wu06}  Wu, J., Zhou, X., Wu, X.-B., 
Liu, F.-K., Peng, B., Ma, J., Wu, Z., Jiang, Z., Chen, J., 2006, \aj, 132, 1256. 
\bibitem[Zeldovich (1964)]{zel64} Zeldovich, Ya.B. 1964, Sov.Phys.Doklady, 9, 195 
\end{thebibliography}
\end{document}